\documentclass[prl,showpacs,preprintnumbers,amsmath,amssymb,twocolumn]{revtex4}
\usepackage{amsmath}
\usepackage{amsthm}
\usepackage{amssymb}
\usepackage{graphics}

\newcommand{\be}{\begin{equation}}
\newcommand{\ee}{\end{equation}}

\newcommand{\Om}{\ensuremath{\Omega}}

\newcommand{\Lap}{\ensuremath{\Delta }}

\newtheorem{prop}{Proposition}

\usepackage{bm}

\begin{document}

\title{On Information Theory, Spectral Geometry and Quantum Gravity}
\author{Achim Kempf, Robert Martin}
\affiliation{
Departments of Applied Mathematics and Physics\\ University of Waterloo\\
Waterloo, Ontario N2L 3G1, Canada}
\begin{abstract}

We show that there exists a deep link between the two disciplines of
information theory and spectral geometry. This allows us to obtain new
results on a well known quantum gravity motivated natural ultraviolet
 cutoff which describes an upper bound on the spatial density of
information. Concretely, we show that, together with an infrared cutoff,
this natural ultraviolet cutoff beautifully reduces the path integral of
quantum field theory on curved space to a finite number of ordinary
integrations. We then show, in particular, that the subsequent removal of
the infrared cutoff is safe.

\end{abstract}

\pacs{04.60.-m, 04.62.+v, 03.67.-a, 04.60.Pp}

\maketitle Well-known quantum gravity arguments indicate the existence of
a natural ultraviolet (UV) cutoff in nature. In this context, there is
much debate as to whether spacetime is fundamentally discrete or
continuous. Spacetime discreteness would naturally regularize quantum
field theoretic UV divergencies but general relativity naturally lives on
a differentiable spacetime manifold. As was first pointed out in
\cite{ak-firstsampling}, the presence of an information theoretic natural
UV cutoff would allow spacetime to be in a certain sense both discrete and
continuous: spacetime would be described as differentiable manifold while
physical fields possess a merely finite density of degrees of freedom. In
this scenario, when a field is known on an arbitrary discrete lattice of
points whose spacing is at least as tight as some finite value, e.g., at
the Planck scale, then the field is reconstructible at all points of the
manifold. In this way, actions, fields and their equations of motion can
be written as living on a smooth spacetime manifold, displaying, for
example, symmetries such as Killing vector fields, while, completely
equivalently, the same theory can also be written on any sufficiently
dense lattice, thereby displaying its UV finiteness. Continuous external
symmetries such as Killing vector fields are not broken in this scenario
because there is no preference among the lattices of sufficient proper
density. This type of natural UV cutoff could be a fundamental property of
spacetime or it could be an effective description of an underlying
structure within a quantum gravity theory such as string theory or loop
quantum gravity. Indeed, this type of natural UV cutoff has been shown to
arise, see \cite{ak-firstsampling}, from generalized uncertainty relations
of string theory and  general studies of quantum gravity \cite{ucrs}.

The mathematics of continuous functions which can be reconstructed from
their sample values $\{f(t_n)\}$ on any discrete set of points $\{t_n\}$
of sufficiently tight spacing is a well-developed field, called sampling
theory, and it plays a central role in information theory. Shannon
introduced sampling theory in his seminal work \cite{shannon} as the link
between discrete and continuous representations of information. For
example, the basic Shannon sampling theorem applies to functions, $f$,
which possess only frequencies below some finite bandwidth $\Omega$, i.e.,
which are bandlimited. The theorem states that it suffices to know the
discrete values $\{f(t_n)\}$ at a set of equidistant points $t_n$ whose
spacing $t_{n+1}-t_n$ is smaller than $1/2\Omega$ to be able to {\it
perfectly} reconstruct $f(t)$ for all $t$:
\be
f(t) = \sum _{n\in \mathbb{Z}} f(t_n) \frac{\sin \left( \Om (t-t_n)
\right)}{ \Om (t-t_n)} \label{eq:shannon}
\ee
Sampling theory for functions of one variable is a mature field,
\cite{benedetto}, with countless applications from communication
engineering to pure mathematics.
Sampling theory is much less developed for the case of bandlimited
functions in $R^n$, though Landau \cite{landau} established the average
spacing which sample points $\{t_n\}$ must have if the sample values
$\{f(t_n)\}$ are to allow the stable reconstruction of bandlimited
functions $f(t)$ for all $t\in R^n$. Of interest for our purposes is of
course the generalization of sampling theory for functions on generic
non-compact curved spaces, a field that is so far only at its beginning,
\cite{pesenson,kempf04}.

In this Letter, we find a powerful new tool for developing sampling theory
on generic non-compact curved spaces, namely  a deep relationship between
sampling theory and spectral geometry. In particular, we use the new tool
to derive results on quantum field theory (QFT) on generic non-compact
curved spaces with the sampling theoretic UV cutoff. Our aim is to model
the behavior of QFT, on a curved background spacetime, as the Planck scale
is approached from lower energies.

We begin by considering the path integral formulation of euclidean
signature QFT. The spectrum of the Laplacian is an invariant of the
manifold and therefore cutting off the spectrum of the Laplacian, say
close to the Planck scale, is to covariantly project the Hilbert space of
square integrable functions down to a Hilbert space of bandlimited
functions. This UV cutoff could arise in various ways depending on the
underlying theory of quantum gravity. For example, the full effective
action could contain a power series in the Laplacian with a finite radius
of convergence, see \cite{kempf04}. The subspace of covariantly
$\Om-$bandlimited functions on a curved manifold $M$ is then defined as
$B(M, \Om) := P_{[0, \Om ^2 ]} (-\Lap) L^2 (M)$. Here, $P_{[0, \Om ^2
]}(-\Lap)$ is the projector onto the subspace spanned by the
eigenfunctions to $-\Lap$ whose eigenvalues lie in the interval $[0, \Om
^2 ]$. To be precise, using the functional calculus for self-adjoint
operators, $P_{[0, \Om ^2 ]}(-\Lap)$ is the characteristic function of the
interval $[0, \Om ^2]$ with the Laplacian as its argument.

While, therefore, it is straightforward to define bandlimitation
covariantly, the fundamental problem is to show that these bandlimited
functions can be reconstructed from their discrete samples if those
samples are taken at a suitable average spacing that is finite. Our
strategy for developing sampling theory on noncompact curved spaces is to
reconsider the very simplest instances of sampling theory and to build up
the case of sampling on generic noncompact curved spaces from there. To
this end, consider the simple case of an $N$-dimensional function space,
$F$, spanned by some generic basis functions $\{b_i(x)\}_{i=1...N}$, i.e.,
all $f$ obey $f(x)=\sum_{i=1}^N\lambda_i ~b_i(x)$ for some
$\{\lambda_i\}$. There automatically holds a sampling theorem for this
function space: assume we know of a function $f\in F$ only its amplitudes
$a_n=f(x_n)$, for $n=1...N$ at some $N$ generically chosen points $x_n$,
i.e.,
\begin{equation}
f(x_n)=a_n=\sum_{i=1}^N \lambda_i ~b_i(x_n)\label{e1}
\end{equation}
Then, Eq.\ref{e1} generally allows us to determine the coefficients
$\lambda_i$ and therefore $f(x)$ for all $x$. This is because for generic
basis functions $b_i$ and sample points $x_n$ the $N\times N$ matrix
$B=(b_i(x_n))_{i,n=1...N}$ has a nonvanishing determinant and is therefore
invertible, so that we obtain: $\lambda_i=\sum_{j=1}^N B^{-1}_{ij} ~a_j$
and therefore
$$
f(x) = \sum_{n=1}^N f(x_n) G(x_n,x) \mbox{~~~~for all}~x
$$
where the reconstruction kernel $G$ reads: $G(x_n,x)=\sum_{i=1}^N
B^{-1}_{ni}~b_i(x)$. In practice, we are interested in sampling theory for
infinite-dimensional function spaces. As we therefore let
$N\rightarrow\infty$ the number of basis functions and correspondingly
also the necessary number of sample points diverges. It depends crucially
on the particulars of the set of basis functions whether or not for
$N\rightarrow\infty$ these infinitely many sample points can still be
chosen at a finite spacing, i.e., whether or not there is a sampling
theorem for $N\rightarrow \infty$. Our aim now is to pursue this analysis
on generic noncompact Riemannian manifolds.

Our starting point is the fact that the space of bandlimited functions
 on any Riemannian manifold $K$ is as simple as described
above, namely finite dimensional, if the manifold is compact: it is known,
see \cite{davies}, that, e.g., for Dirichlet and v. Neumann boundary
conditions, the Laplacian $\Lap _K $ on $K$ has only purely discrete
eigenvalues, of finite multiplicity $ 0 \leq \lambda _1 < \lambda _2 < ...
\rightarrow \infty$ and without finite accumulation points. Thus, imposing
a bandlimit by cutting off the spectrum of the Laplacian on $K$ implies
that the space of bandlimited functions on $K$ is in fact finite
dimensional. Thus, the euclidean path integral on $K$ consists of finitely
many ordinary integrations.

For later reference we note a subtle point: the space of $\Om-$bandlimited
functions on $K$, defined as $B(K, \Om) := P _{[0, \Om ^2]} (-\Lap _K) L^2
(K)$, is a subspace in $L^2 (M)$ through $B(K,\Om) = P _K P _{[0 ,\Om ^2 ]
} (-\Lap _K ) P _K L^2 (M)$. Here, $P _K$ is uniquely the projector of
$L^2 (M)$ onto $L^2 (K)$ but $P _{[0 ,\Om ^2 ] } (-\Lap _K )$ depends on
the choice of self-adjoint extension of $\Lap _K$, i.e., on the choice of
boundary conditions on $K$, a dependence which we will have to control.

Now our strategy for developing sampling theory on a non-compact
Riemannian manifold $M$ is to choose a sequence of nested compact
sub-manifolds $K_i$, with the same dimension as $M$, such that their union
is all of $M$. Physically speaking, we are imposing and removing an
infrared (IR) cutoff. We know that the space of band\-limited functions on
$M$ is infinite dimensional, i.e., as we consider larger and larger
portions $K_i$ of $M$, the dimension $N$ of the space of functions over
$K_i$ with the same bandlimit (say at the Plank scale) should grow and
eventually diverge. (Thereby, the path integral will again become an
infinite number of integrations.) The key question is how quickly the
dimension $N$ diverges as a function of the size of $K_i$. The speed of
this growth determines if the sample density can be kept finite as the
size of the portions $K_i$ of $M$ diverges. Here, we observe a deep
connection to spectral geometry:

Spectral geometry studies the relationship between the size and
shape of a Riemannian manifold and the spectra of differential
operators, such as the Laplacian on that manifold. See, e.g.,
\cite{davies} or the well-known article ``Can one hear the shape
of a drum?" by Kac \cite{kac}. Spectral geometry therefore
naturally encompasses the mathematics of general relativity, i.e.,
differential geometry, as well as that of quantum theory, i.e.,
functional analysis. Spectral geometry has found various
applications from number theory to noncommutative geometry
\cite{connes}. Of interest for us here is a fundamental result of
spectral geometry called Weyl's asymptotic formula for compact
Riemannian manifolds: it states that the number, $N(\Omega,K)$, of
the Laplacian's eigenvalues below some value $\Omega ^2$
approaches
\be N(\Omega, K) \rightarrow \frac{\Omega^d V(B_d)
V(K)}{(2\pi)^d}\label{weyl} \ee for large $\Om$, where $V(K)$ is
the volume of $K$ and $V(B_d)$ is the volume of the
$d-$dimensional unit ball in $\mathbb{R} ^d$. For intuition,
consider the special case of the Hamiltonian, $H=-\partial^2$, of
the free particle on an interval of length $L$: the spectrum of
$H$ is proportional to $(n/L)^2$ and therefore the number
$N(\Omega)$ of eigenvalues of $H$ that are below some cutoff value
$\Omega^2$ is proportional to the size $L$ of the interval:
$N(\Omega,L)\propto\Omega L$. The same argument applies in
$\mathbb{R}^d$. Weyl's asymptotic formula shows that the spacing
of eigenvalues behaves for large $\Omega$ the same way also on
curved manifolds. This is plausible since modes of large enough
eigenvalues possess wavelengths that are short compared to the
smallest length scale of the curvature.

Equation (\ref{weyl}) provides exactly the information that we need, at
least in the physically relevant case where the curvature and the UV
cutoff are such that we are in the asymptotic regime where Weyl's formula
holds. Intuitively, this is the case where the UV cutoff length scale is
small compared to the smallest curvature length scale of the manifold. We
will consider the alternate case at the end. Now Weyl's asymptotic formula
shows that the dimension, $N$, of the space of bandlimited functions on
$K_i$ is proportional to the volume $V(K_i)$ of the compact submanifolds
$K_i$. Thus, for larger and larger $K_i$, the spatial density
$N(K_i)/V(K_i)$ of sample points does in fact remain finite, at
$(\Omega/2\pi)^d$. Note that for flat space this agrees with Landau's
value, as it should. The use of Weyl's formula on successively larger
submanifolds of a generic non-compact curved manifold, i.e., while
imposing and removing an IR cutoff, therefore provides us with a powerful
new tool both for sampling theory and for QFT. On one hand, Weyl's formula
should be very useful for further developing sampling theory on curved
space, e.g., to investigate when a particular set of sample points is
dense enough to allow the reconstruction of functions from those sample
values. On known results, see \cite{pesenson}. On the other hand, the
controlled imposing and removing of the IR cutoff in curved space, with
the UV cutoff, can render the path integral finite and well-defined. In
particular, because of the sampling property, the fields can be viewed as
living on a differentiable manifold and also as living on any one of a set
of sufficiently dense lattices, as discussed above. It is straightforward,
for example, to calculate the representation of the derivatives on the
lattice representations, as will be shown explicitly in a follow-up paper.
While this approach so far only applies to the euclidean signature, it
does cover the case of Wick-rotated spacetimes and the case of space-like
hypersurfaces (in particular, if the d'Alembertian possesses an elliptic
spatial part). It could therefore provide a covariant quantitative
framework for holography.

At this point we need to consider, however, that, in the presence of a UV
cutoff, it is in fact non-trivial to keep the imposition and subsequent
removal of an IR cutoff under control. In the path integral of quantum
field theory an IR cutoff on a curved manifold $M$ is the restriction of
the space of fields $L^2(M)$ to $L^2(K)$ where $K\subset M$ is a
submanifold with compact closure. Removing the IR cutoff corresponds to
considering a nested sequence of ever larger submanifolds $K_i$ (whose
closures are compact), and whose union is all of $M$. One possibility
would be to impose the UV cutoff first, i.e., to restrict the space of
fields that is being integrated over in the path integral to
$B(M,\Omega)$, then to impose an IR cutoff, perform calculations, and
finally remove the IR cutoff. Technically, we would work with the image of
$L^2 (M)$ under the operators $P_n P_\Om$ and then take the limit as
$n\rightarrow \infty$, where $P_\Om$ projects onto $B(M,\Om)$ and $P_n$
projects onto $L^2 (K_n)$. This procedure, however, is not practical.

First we notice that the operator $P_n P_\Om$ is not a projector because
$P_n$ and $P_\Om$ do not commute. In fact, the range of $P_n P_\Om$ is not
closed and is therefore not the image of $L^2 (M)$ under any projector. In
the path integral it would not be straightforward to restrict the fields
to this subspace so that the UV cutoff on the full manifold is regained as
the IR cutoff is removed. In fact, the subspace resulting from imposing
first the UV and then the IR cutoff does not obey an UV cutoff on $K_n$,
i.e., performing the path integral will be no simpler than with no UV
cutoff on the full manifold. The reason can be traced to the existence of
superoscillations \cite{pollak,ferreira-kempf}: even for the simple case
where $M$ is the real line, it is known that the space of
$\Omega$-bandlimited functions contains functions that oscillate
arbitrarily fast on any given finite interval. This means that the
projection of $B(\mathbb{R} , \Om)$ onto the space of functions with
support on a finite interval $I$ does not yield a space $B(I, \Om)$ of
bandlimited functions on that finite interval. Instead, it yields a
subspace which is dense in $L^2 (I)$ \cite{pollak}, implying that imposing
first an UV and then an IR cutoff yields an infinite dimensional subspace
of functions even on the compact submanifolds $K_n$.

Instead, as we will now show, it is practical to first restrict the fields
to $L^2 (K_n)$ and then to cut off the spectrum of the Laplacian on $K_n$,
namely, to project $L^2(M)$ with the projector $P_{K_n , \Om} P_n = P_n
P_{K_n ,\Om} P_n$ where $P_{K_n , \Om } = P_n P_{[0, \Om ^2]} (\Lap _n)
P_n$ projects onto $B(K_n ,\Om)$. This is what we did above and we know
that the resulting space of fields, $B(K_n ,\Om )$ is a closed,
finite-dimensional subspace, so that the path integral in the presence of
both the UV and IR cutoffs is then simple, well defined and has the
sampling property. We have to show, however, that the removal of the IR
cutoff is under control, i.e., that one recovers the full theory as
$n\rightarrow \infty$.

To this end, we need to consider the functionals on fields $\phi$, such as
the action functional $S[\phi]$ in the path integral formulation or such
as the state functionals $\Psi[\phi]$ in the Schr\"odinger formulation of
QFT. We have to show that the evaluation of such functionals in the full
(i.e. only UV cutoff) theory agrees with the limit of evaluating these
functionals on successively larger submanifolds. Concretely, if $\Psi $ is
such a functional then in order that the removal of the IR cutoff be safe
we need that
\be \Psi [ P_{K_n , \Om} \phi ] \rightarrow \Psi [ P_\Om \phi ]
\label{eq:irgone} \ee as $n\rightarrow \infty$ for any $\phi \in
L^2(M)$. For any continuous $\Psi$, equation (\ref{eq:irgone})
will hold provided that $P_{K_n, \Om} \phi \rightarrow P_\Om  \phi
$ for all $\phi \in L^2 (M)$, \emph{i.e.}, provided that we can
show that $P_{K_n , \Om}$ converges strongly to $P_\Om$. We claim
that, in spite of the above-discussed superoscillations and in
spite of the non-uniqueness of the boundary conditions (as well as
self-adjoint extensions, eigenvectors and spectra) of the
IR-cutoff Laplacians the following holds:

\begin{prop}
 The projector $P_\Om$ is the strong limit of the sequence of
 projectors $\{P_{K_n ,\Om}\}_n$, i.e., $P_{K_n ,\Om}
 \stackrel{s}{\rightarrow} P_\Om$.
\end{prop}
Here, $``\stackrel{s}{\rightarrow}"$ denotes convergence in the strong
operator topology, \cite{reed}. The detailed proof is somewhat technical
and lengthy and we therefore postpone its full presentation to the
follow-up paper. Let us here however outline the method that we use for
the proof. First, let $M$ be any $C^{\infty}$ complete Riemannian manifold
and let $\{K_n\}$ be a sequence of open submanifolds $K_n \subset M$ which
have compact closures and which are nested $K_n \subset K_{n+1}$ and which
cover all of $M$, $\bigcup _n K_n =M$.
Then, let $\Lap _n$ be any self-adjoint extension of the Laplacian on
$K_n$. The Laplacians are unbounded operators and our first aim is to
show, therefore, that for any bounded continuous function $g:
\mathbb{R}\rightarrow \mathbb{R}$ we have that $g(\Delta_n)
\stackrel{s}{\rightarrow} g(\Delta)$. To this end, we prove a stronger
proposition:
\begin{prop}
    The Laplacian of $M$ is the strong graph limit (and, equivalently, the strong
resolvent limit) of any sequence of self-adjoint Laplacians $\Lap _n $ on
the compact submanifolds $K_n$, i.e., $\Lap _n \stackrel{sg}{\rightarrow}
\Lap$. \label{prop:sglim}
\end{prop}
In the literature, special cases for one dimension and for flat space are
known, see \cite{weidmann,stollmann,martin}. Useful for our purpose here
is the known result that if $A_n \stackrel{sg}{\rightarrow} A$ and if $a,b
\in \mathbb{R}$,~ $a<b$ and $a,b$ are not eigenvalues of $A$ then $P
_{(a,b)} (A_n)$ converges strongly to $P_{[a,b]} (A)$, \cite{reed}.
Therefore, to prove proposition 1, namely that $P_{[0 , \Om ^2]} (\Lap
_n)$ converges strongly to $P_\Om$, we show that $\Lap _n
\stackrel{sg}{\rightarrow} \Lap$. (We can assume that $0, \Om ^2$ are not
eigenvalues of $\Lap$, if need be by a suitable arbitrarily small change
of the spectral interval that we project on.) With this result it is then
straightforward to show that $P_{K_n , \Om } = P_n P_{[0, \Om ^2]} (\Lap
_n) P_n $ also converges strongly to $P_\Om$. In fact, the exact same
proof also goes through for pseudo-Riemannian manifolds $M$ for which the
Dirac operator $D$, or the d'Alembertian, $\Box$, is essentially
self-adjoint on $C_0 ^{\infty} (M)$. That is, if $K_n$ is a sequence of
nested compact submanifolds of $M$ as before then $P_n P _{(a,b)} (\Box _n
) P_n $ converges strongly to $P_{[a,b]} (\Box)$ provided again that $a,b$
are not eigenvalues of $\Box$. The present work therefore lays the
groundwork also for generalizing sampling theory to Lorentzian manifolds.

Let us now recall that so far we restricted attention to QFT on fixed
background spacetimes which are curved only on length scales that are
significantly larger than the UV cutoff scale. The alternate case where
spacetime is curved on scales that are significantly smaller than the UV
cutoff scale for particle fields is physically likely excluded by a
natural UV cutoff on curvature, at about the same scale. The remaining
case of a manifold with bounded curvature, say a bounded geometry, which
possesses curvature at length scales down to close to the UV cutoff scale
is clearly of interest and points toward the need to develop sampling
theory for manifolds themselves, with a suitable UV cutoff for curvature
then playing the role of the bandwidth. The problem of imposing a
curvature cutoff is nontrivial, especially on Lorentzian manifolds
\cite{schuller}. Sampling theory of geometry should then very
interestingly intertwine, roughly speaking, the location and the amplitude
of samples of the metric field.

The present sampling theoretic methods could be of interest in any theory
of quantum gravity. For example, in the context of spin foam models,
\cite{qg}, the lattice representations of sampling theory could provide
discretizations of the manifold which automatically preserve the dimension
and topology of the manifold. The sampling theoretic UV cutoff has already
been implemented in inflationary cosmology, \cite{cosm}, and it is
therefore linked to a potential experimental window to Planck scale
physics.

\end{document}